\documentclass[a4paper,twoside,reqno]{bjp}
\usepackage{graphicx}
\usepackage{cite}
\usepackage{amssymb,amsmath,amscd,amsthm}
\usepackage{times}

\usepackage[bookmarks=false]{hyperref}
\hypersetup{%
    colorlinks=true,        
    linkcolor=blue,          
    citecolor=blue,         
    urlcolor=blue           
    }

\usepackage{geometry}
\allowdisplaybreaks

\begin{document}
\def\bb    #1{\hbox{\boldmath${#1}$}}
 \def\rr    {{R}_0 \!\!\!\!\!{}^{{}^{{}^{\circ}}}\,}  
\def\RR         {R\!\!\!{{}^{{}^{{}^{\circ}}}}}~

\title{Toroidal States of the $^{12}$C  
Nucleus\,\thanks{Talk presented at the International Workshop on Shapes and Dynamics of Atomic Nuclei at the Bulgarian Academy of Sciences, Sofia, Bulgaria, 3-5 October, 2019.  }}

\runningheads{Cheuk-Yin Wong~~~~~}{Toroidal States of  $^{12}$C ~~~~~~~~~~}

\begin{start}{
\author{Cheuk-Yin Wong}{ }

\address{Physics Division, Oak Ridge National Laboratory, Oak Ridge, TN 37831 USA 
}{  }

\received{Day Month Year (Insert date of submission)}
}

\begin{Abstract}

Among the states of $^{12}$C, there is an important  subset of $K$=0 and $K$=$I$  planar intrinsic  states in which the intrinsic motion of the nucleons are confined in the planar region defined by the three-alpha cluster or by  their generated toroid.   The intrinsic nuclear densities of these states are toroidal in nature.  We study these $^{12}$C toroidal states from the generator-coordinate viewpoints in both the alpha cluster model and the  toroidal shell model.  Numerical solutions in the toroidal mean field approximation  are examined to pave the way for future extensions and refinements.

\end{Abstract}

\begin{KEY}
 $^{12}$C Hoyle state,  toroidal $^{12}$C  states, generator coordinate method
\end{KEY}
\end{start}


\section{Introduction:}

John Wheeler suggested that under an ``extreme behavior", the nuclear fluid
may assume a toroidal shape.  He encouraged his students and fellow physicists to think about  where such extreme  behavior might occur and to look for it \cite{Whe50,Whe98}.  
Possible extreme behavior might occur in the presence of  (i) a large Coulomb energy \cite{Whe50,Won73}, (ii) a strong nuclear shell effect \cite{Won73}, and/or (iii) a large angular momentum \cite{Won78}.   There is a strong shell effect  of   a doubly closed-shell  for $^{12}$C  in a
toroidal potential and the extrapolation from heavier nuclei points to a possible low-lying toroidal state in   $^{12}$C \cite{Won73},  suggesting  that  the $^{12}$C nucleus may be a favorable candidate for a toroidal configuration \cite{Won19}.
The study of light toroidal nuclei gains additional impetus recently because the effects of strong toroidal shells and large angular momenta lead to toroidal high-spin isomers in many light alpha-conjugate nuclei, as shown 
in different theoretical calculations \cite{Ich12,Sta14,Sta15,Sta15a,Sta16,Cao19} and 
 in possible experimental observations of highly excited states in $^{28}$Si.

For the $^{12}$C nucleus in particular, 
 Wheeler's  model of a triangular  3$\alpha$ cluster
\cite{Whe37} has been studied extensively
\cite{Whe37a,Hil53,Gri57,Kam81,Ueg79,Kan07,Che07,AR07}.   However, Wheeler's other concept of a possible toroidal nucleus
\cite{Whe50,Whe98} has up to now not been applied to the $^{12}$C
nucleus.  Wheeler's two different
concepts should play their separate and important roles under
different probes of the nucleus.  In matters of 3$\alpha$ decay and
the escape through the external Coulomb barrier, the 3$\alpha$ cluster
description is clearly the simpler description.  However, because 
 nucleons can traverse azimuthal orbitals in a
toroidal nucleus with low energies and can be excited to higher orbitals with low excitation energies,  a description in terms of a
$^{12}$C nucleus in toroidal doubly-closed shells may be an efficient
description in matters associated with particle-hole excitations and
in the density of  particle-hole multiplet states.   
Furthermore, the
toroidal concept provides a novel geometrical insight, organizes
useful correlations, helps guide our intuition, and may find many
applications involving the $^{12}$C nucleus.  
It is therefore useful to develop the toroidal concepts for the $^{12}$C nucleus.
 
\section{ Generator Coordinate Equation for the Planar Alpha Cluster Model}

We envisage that among the states of $^{12}$C, there is an important  subset of oblate $K$=0 and $K$=$I$ planar intrinsic  states in which the intrinsic motion of the nucleons are confined in the planar region  defined by the cluster of three alpha particle or by the toroid generated by the planar rotation of the 3$\alpha$ cluster.   States in this subset are characterized by the angular momentum component quantum number $K$ normal to the plane, and appear as the band head of collective rotational states when the states are  projected into the full three-dimensional space.
States of this kind can be described in many different ways, depending on an additional assumption on how the intrinsic shape is generated.  For example, we can follow Wheeler \cite{Whe37} and assume the intrinsic shape to be generated by an even more basic  triangular cluster of three alphas particles, $|\phi_j (\Delta)\rangle$, and consider as in
Griffin, Hill, and Wheeler \cite{Gri57,Hil53} a trial 
wave function $|\Psi^K\rangle$ as a coherent sum 
 of these antisymterized triangular three-alpha  cluster  wave functions $|\phi_j (\Delta)\rangle$ of different triangular orientations on the plane
\begin{eqnarray}
|\Psi^K\rangle=\int  d\gamma ~e^{-i K \gamma}~e^{i\gamma {\hat L}_z} \sum_j f_j^K |\phi_j (\Delta)\rangle.
\end{eqnarray}
Here,  $\hat L_z$ is the rotation  operator about the body-fixed $z$-axis on the triangular plane which does not involve the spin contribution,
 $\gamma$ is the corresponding Euler angle for such a rotation,
 and $f_j^K$ are the Griffin-Hill-Wheeler generator coordinate amplitude for the state 
$|\Psi^K\rangle$, and the operator 
$\int d\gamma e^{-i K \gamma}~e^{i\gamma {\hat L}_z} $ projects out states of good $K$ quantum number.
The generator coordinate  projection sum of the cluster wave function
$|\phi_i (\Delta)\rangle$
 on the triangular plane over all the orientations specified by $\gamma$ leads to wave functions with toroidal characteristics, which we can label as $|\phi_j^K({\rm toroid})\rangle$,
\begin{eqnarray}
|\phi_j^K({\rm toroid})\rangle=\int  d\gamma ~e^{-i K \gamma}~e^{i\gamma {\hat L}_z} |\phi_j (\Delta)\rangle.
\label{mix1}
\end{eqnarray}
In terms of these toroidal wave functions $ |\phi_i^K ({\rm toroid})\rangle$ projected from the basic three-alpha cluster state $|\phi_i (\Delta)\rangle$, the trial wave function becomes
\begin{eqnarray}
| \Psi^K\rangle=\sum_i f_j^K  |\phi_j^K({\rm toroid})\rangle.
\label{mix2}
\end{eqnarray} 
Quantization of the system can be carried out by minimizing the energy with respect to the  variation of the trial wave function $|\Psi^K\rangle$ under the constraint of a fixed normalization,  
resulting in the Griffin-Hill-Wheeler equation for  $f_j^K$,
\begin{subequations}
\begin{eqnarray}
&&\sum_j \bigl [ H_{ij}   - E  B_{ij}\bigr ]  f_j^K =0,
\label{ghw}
\\
\text{where~~~~~~~~}~~~~~~~~~~&&H_{ij}= \langle  \phi_i^K ({\rm toroid}) |H|  \phi_j^K ({\rm toroid}) \rangle,~~~~~~~~~~~~~~~~~~~~~~~~~~~~~~~
\label{hij}
\\
&&B_{ij}= \langle  \phi_i^K ({\rm toroid}) | \phi_j^K ({\rm toroid}) \rangle.
\label{bij}
\end{eqnarray} 
\end{subequations}
The generator-coordinate method involves introducing a two-body interaction, solving  the above Griffin-Hill-Wheeler, and projecting  the intrinsic  toroidal \break  solution onto the full three-dimensional laboratory space  for states with angular momentum quantum numbers $I$ and $M$ afterwards.

\section{  Generator Coordinate Equation for the Toroidal Shell Model}

The expansion  of the physical wave function $|\Psi^K\rangle$  in Eq.\ (\ref{mix1}) in terms of the 3$\alpha$ cluster 
wave functions $|\phi_j(\Delta)\rangle$ of Eq.\  (\ref{mix2})  
 is useful in problems where the cluster properties  manifest
themselves.  However,  it   
is
not the only way to construct a trial wave function $|\Psi^K\rangle$.  
Because 
 nucleons can traverse azimuthal orbitals in a
toroidal nucleus  with low energies and 
they can be scattered from good-Lambda orbitals just
below the Fermi level to orbitals just above the Fermi level
with low excitation energies,  a description in terms of a
$^{12}$C nucleus in toroidal configurations  may be an efficient
description in matters associated with particle-hole excitations.  
For such problems, the presence of the toroidal degree of freedom allows an alternative  trial wave function for the state with quantum number $K$,
\begin{eqnarray}
 |\Psi^K\rangle=\sum_j g_j^K  |\Phi_j^K({\rm toroid})\rangle,
\label{mix3}
\end{eqnarray}
where $|\Phi_j^K({\rm toroid})\rangle$ is chosen to be those obtained in a  toroidal shell model, either self-consistently  from a mean-field approximation or non-self-consistently from a single-particle model with an assumed
toroidal shape.    The Griffin-Hill-Wheeler equation for the amplitude $g_j^K$ is
\begin{subequations}
\begin{eqnarray}
&&\sum_j \bigl [ H_{ij}   - E  B_{ij}\bigr ]  g_j^K =0,
\label{ghw}
\\
\text{where~~~~} ~~~~~~~~~~~~&&H_{ij}= \langle  \Phi_i^K ({\rm toroid}) |H|  \Phi_j^K ({\rm toroid}) \rangle,
~~~~~~~~~~~~~~~~~~~~~~~~~~~~~~~
\label{hij2}
\\
&&B_{ij}= \langle  \Phi_i^K ({\rm toroid}) | \Phi_j^K ({\rm toroid}) \rangle.
\label{bij2}
\end{eqnarray} 
\end{subequations}
The quantities $H_{ij}$ and $B_{ij}$ in 
(\ref{hij2}) and (\ref{bij2}) are now defined in terms of $|\Phi_j^K\rangle$ from a toroidal shell model in place of $|\phi_j^K\rangle$ from the projection of three-alpha cluster states.  
Different choices of the trial wave functions from either 
Eq.\ (\ref{mix1}) (and (\ref{mix2}))  or Eq.\  (\ref{mix3}) 
should converge to the same physical state $ |\Psi^K\rangle$, provided they have been chosen efficiently with a sufficient number of components in the expansion.

\section{Toroidal Mean Field Approximation}

\begin{figure} [h]
\includegraphics[scale=0.87]{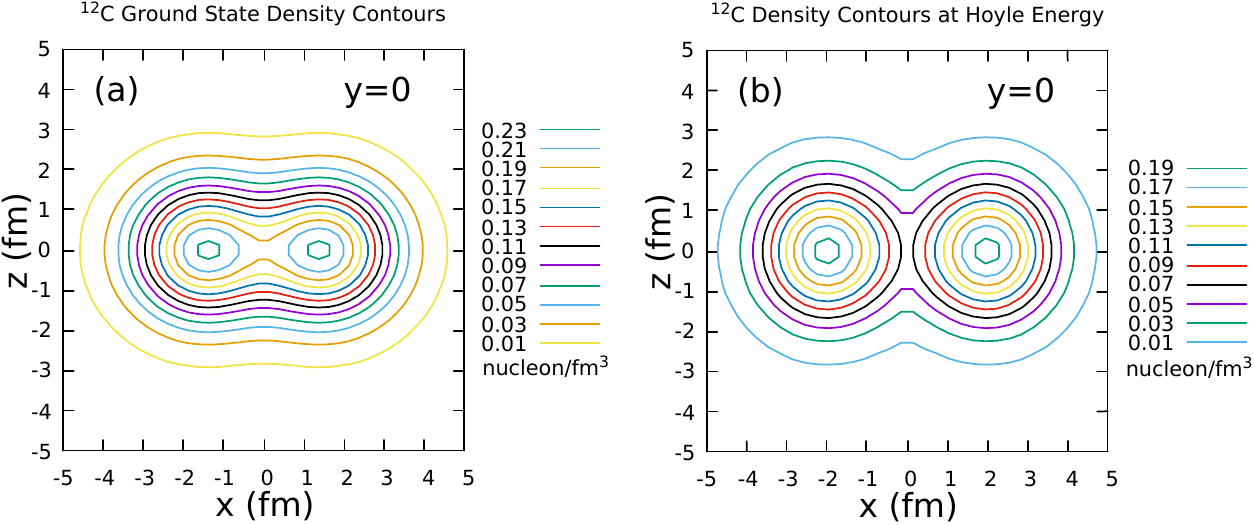}
\centering
%
\caption{(color online). The nuclear density  of the $^{12}$C on the $y=0$ plane  for (a) ground state and (b) at the Hoyle energy, obtained with the 
Skyrme SkM$^*$ interaction in the mean-field approximation.}
\label{twoden}
\end{figure}

The simplest expansion in Eq.\ (\ref{mix3}) consists of a single term $|\Phi_{i=1}^K\rangle$ 
 in
the mean-field approximation 
in which   we  describe the $^{12}$C nucleus
with a single Slater determinant that takes into account the dominant correlations
with a mean-field interaction. 
Consider as an example the case of $K$=0. 
With an intrinsic axial symmetry about the $z$-axis, the $^{12}$C nucleus  with $K$=0 can be represented by the single Slater determinant of neutrons
and protons occupying  the lowest single-particle 
$|\Lambda, \Omega_z\rangle$ states, 
where $ \Lambda$=$|\Lambda_z|$, $\Lambda_z$ is the $z$-component of the orbital angular momentum, $\Omega_z$=$\Lambda_z+s_z$, and $s_z$ is the spin component  along the $z$ axis.   Limiting our attention on the state with the lowest oscillation quanta in the $\rho$- and $z$- directions for $^{12}$C, the occupied  single-particles states are $|0,\pm 1/2\rangle$,
$|1,\pm 3/2\rangle$, and $|1,\pm 1/2\rangle$. 
Assuming the same set of wave functions for neutrons and protons and
neglecting the spin-orbit interaction, we write down 
 the variational spatial wave functions of the occupied
single-particle states in
terms of variational parameters $(R,d, a_2)$
\begin{subequations}
{
\begin{eqnarray}
&&\Psi_{\Lambda_z \Omega_z}(\rho,z,\phi)={\cal R}_{\Lambda}(\rho )Z(z) [\Phi_{\Lambda_z} (\phi)\chi_{s_z}]^{\Omega_z},
\label{Psi}
 \\
\text{where~}~~~ && {\cal R}_{\Lambda} (\rho)=N_{\Lambda}\rho^{\Lambda}
\exp \left \{  -\frac{(\rho-R)^2}{2(d^2 e^{2a_2}/\ln 2)}\right \},~~~~\Lambda=0,1,
\hspace*{1.4cm}
\label{R}
\\
&&~~Z(z)=N_Z \exp \left \{- \frac{ z^2}{2(d^2e^{-2a_2}/\ln 2)}  \right \},
\label{Z}
\\
&&~~\Phi_{\Lambda_z} (\phi)=\frac{e^{i\Lambda_z \phi}}{\sqrt{2\pi}},
\label{Phi}
\end{eqnarray}
}
\label{wave}
\end{subequations}
\hspace*{-0.1cm}with normalization constants $N_Z$ and 
$N_\Lambda$, and $[\Phi_{\Lambda_z} (\phi)\chi_{s_z}]^{\Omega_z}$   to denote the coupling of 
orbital $\Lambda_z$ and spin $s_z$ to $\Omega_z$.
We utilize the density-dependent Skyrme SkM$^*$ interaction \cite{Bar82} that is designed to describe well the surface and bulk properties and large quadrupole deformation properties.  The set of variational parameters $(R$=0, $d$=1.49~fm, $a_2$=$-0.089)$
gives the ground state of the system.   The nuclear density  on the 
$x$-$z$ plane at 
$y$=0 is given in Fig.\ \ref{twoden}. 
The equidensity surfaces in the low density region are
nearly oblate ellipsoids.  When the density increases to $n \ge
0.21$ /fm$^3$, the equidensity surfaces turn into toroids.  
The ground state of the $^{12}$C nucleus  has a dense toroidal 
core immersed in  oblate ellipsoids.

In the mean-field approximation,  we find that the
case of ($R$=1.16 fm, $d$=1.370 fm, $a_2$=-0.007) corresponds to an
excitation energy of the Hoyle energy at $E_x$=7.65 MeV with 
the corresponding nuclear density on the $x$-$z$ plane at $y$=0  shown in
Fig.\ \ref{twoden}(b).  One observes that for
this state at the Hoyle energy, the equidensity surfaces with density $n\ge 0.07$/fm$^3$
appear as separated toroids, while those with lower densities as
spindle toroids.  The meridian cross sections are nearly circular.

 In the mean-field approximation with a single determinant, the ground state with the density shown in Fig.\ \ref{twoden}(a) is a state of lowest energy minimum, but the state at the Hoyle energy
with density shown in Fig.\ \ref{twoden}(b) 
 lies on
an energy slope as a function of $R$, even though it
is stable against variations in $d$ and $a_2$.  It is therefore unstable
against the contraction of the radius parameter $R$ in the mean-field
approximation.  We nonetheless call it a provisional toroidal state
at the Hoyle energy, on account of its toroidal
density shape, pending further investigation of its stability against
$R$ variations beyond the mean field.
Pending modifications beyond the mean field may modify slightly the $R$
location and the shape of the density distribution but will not likely
change its toroidal characteristics.

\section{Phenomenological Study of the $^{12}$C Toroidal Configuration}

In this first exploration of its kind to examine the toroidal degree
of freedom of the $^{12}$C nucleus, it is appropriate to study the
problem from both the  microscopic and phenomenological points of view.
In the phenomenological study as presented in \cite{Won19}, we can search for the signature of the
$^{12}$C nucleus in a toroidal configuration 
so as to facilitate its identification.   The toroidal intrinsic shape 
shows up as a bunching of single-particle
states into ``$\Lambda$-shells" whose spacing is intimately tied to
the size of the toroidal major radius.  This set of single-particle
shells will generate a distinct pattern of particle-hole multiplet
excitations 
between one toroidal single-particle shell to another. 
From such a
signature and experimental data, we find 
phenomenologically
that the Hoyle state and many
of its higher excited states may be tentatively attributed to those of
the $^{12}$C nucleus in a toroidal configuration  \cite{Won19}: 

\begin{itemize}
\item the matching of the gross structure of the low-lying spectrum with the toroidal signature, 

\item the approximate equality  of the strengths  of the  excitation function
in the $^{11}$B($^3$He,d)$^{12}$C$^*$$\to$3$\alpha$  reaction for toroidal states within a multiplet, 

\item  
at low excitation energies, the presence of the underlying structure of unresolved  $^{12}$C states in the $^{11}$B($^3$He,d)$^{12}$C$^*$$\to$3$\alpha$  reaction and the presence of as yet unidentified  members of  the  multiplet states, 

\item at high excitation energies, the presence of a large excitation function  
in $^{10}$B($^3$He,p)$^{12}$C$^*$$\to$3$\alpha$  reactions
and the presence  of a large number of toroidal particle-hole states as a function of energy.

\end{itemize}
There are however many items that need to be further investigated to confirm the presence of such a  toroidal configuration.  
Our suggested description contains future proposed  tests that may be able to
shed more lights on the proper description of the states of $^{12}$C \cite{Won19}.

\section{Residual Octupole-Octupole Interactions in the Mean-Field Approximation}

Even though we do not find a toroidal local energy minimum as a
function of $R$ in the mean-field approximation with a single-determinant
for $K$=0, we are however motivated
to continue the theoretical  search  for an energy minimum in view of the many pieces of experimental
evidence supporting the tentative identification the Hoyle state and
many of its excited states as  toroidal $^{12}$C states, as discussed in \cite{Won19}. From intuitive viewpoints,
we are further encouraged by the small energy separation between the
Hoyle state and its excited states, by the large number of both the
identified and the un-identified broad excited states, by the close average
energy spacing between the states, and by their predominance in their
decay into three alpha particles. These characteristics suggest that
the Hoyle state is intrinsically a spatially extended object that is
capable of possessing a complex particle-hole excitation structure. A local
toroidal energy minimum description is consistent with such a
suggestion. Theoretically we find it promising that the ground state
nuclear density in the mean-field approximation as exhibited in Fig.\ \ref{twoden} already shows a
toroidal structure in its core and the provisional state at the Hoyle
energy in Fig.\ \ref{twoden} exhibits prominent toroidal
characteristics. We also note with interest that
many previous microscopic models with alpha clusters  have been successful in describing both the ground state and the Hoyle state \cite{Kam81,Ueg79,Kan07,Che07,AR07}.  
We are therefore motivated to continue the search for a local energy minimum at the Hoyle energy.

We shall first continue our search within the  limitation of a single determinant,
we would like to explore the effects of residual interactions in the mean-field theory.
We have reasons to infer that the quadrupole  correlations of the nucleons leading to the large quadrupole deformation is properly taken into account by the Skyrme SkM$^*$ interaction.  Indeed,   the toroidal feature of the $^{12}$C
nucleus of the ground state as exhibited in Figs.\ \ref{twoden}(a)
is in agreement with
earlier results  from earlier Hartree-Fock calculations \cite{Rip67}
and the resonating group method
\cite{Kam81}.  In the sense of  the multipole expansion  of the nucleon-nucleon interaction, there  is the possible residual octupole-octupole interaction that may lead to octupole deformation of the triangular type.    We are further guided by results obtained in previous calculations \cite{Kam81,Ueg79,Kan07,Che07,AR07}
where there are cluster solutions at the Hoyle energy and these solutions have prominent triangular cluster characteristics at the Hoyle state.
Accordingly, we postulate the presence of an octupole-octupole interaction in conjunction with the SkM* interaction:
\begin{eqnarray}
V=\frac{1}{2} \sum_i \sum_{j\ne i} {}^{'} V_{{\rm SkM}^*}(i,j) - \frac{\chi}{2} \sum_i  \sum_{j\ne i} {}^{'}Q_{33}(i) Q_{33}(j),
\label{chi}
\end{eqnarray}
where in cylindrical coordinates  
\begin{eqnarray}
Q_{33}(i)=\frac{1}{4}\sqrt{ \frac{35}{4\pi}}(\rho_i/\rr)^3 \cos(3\phi_i),
\end{eqnarray}
$\rr=r_0 A^{1/3}$, and  $r_0$=1.2  fm.  We have included an $1/\rr^3$ scale factor to make the multipole moment $Q_{33}$ dimensionless and $\chi$ in MeV.

To study the additional octupole-octupole interaction, we modify the azimuthal wave function in Eq.\ (\ref{Phi}) to be
\begin{eqnarray}
\Phi_\Lambda (\rho,\phi)=
\frac{[1+ \sigma_3 (\rho/\rr)^3 \cos(3\phi)]e^ {i \Lambda  \phi}}{\sqrt{2\pi}}
\end{eqnarray}
where $\sigma_3$ is the dimensionless  'sausage' deformation parameter of  the $\lambda$=3 order that makes the toroidal nucleus thicker in three sections and thinner in three others  along the toroidal rim  \cite{Won73}.  We can now study the energy surface of $^{12}$C as a function of the variational parameters $(R, d, a_2, \sigma_3)$ for various magnitudes of the octupole-octupole interaction.

\begin{figure} [h]
\includegraphics[scale=0.60]{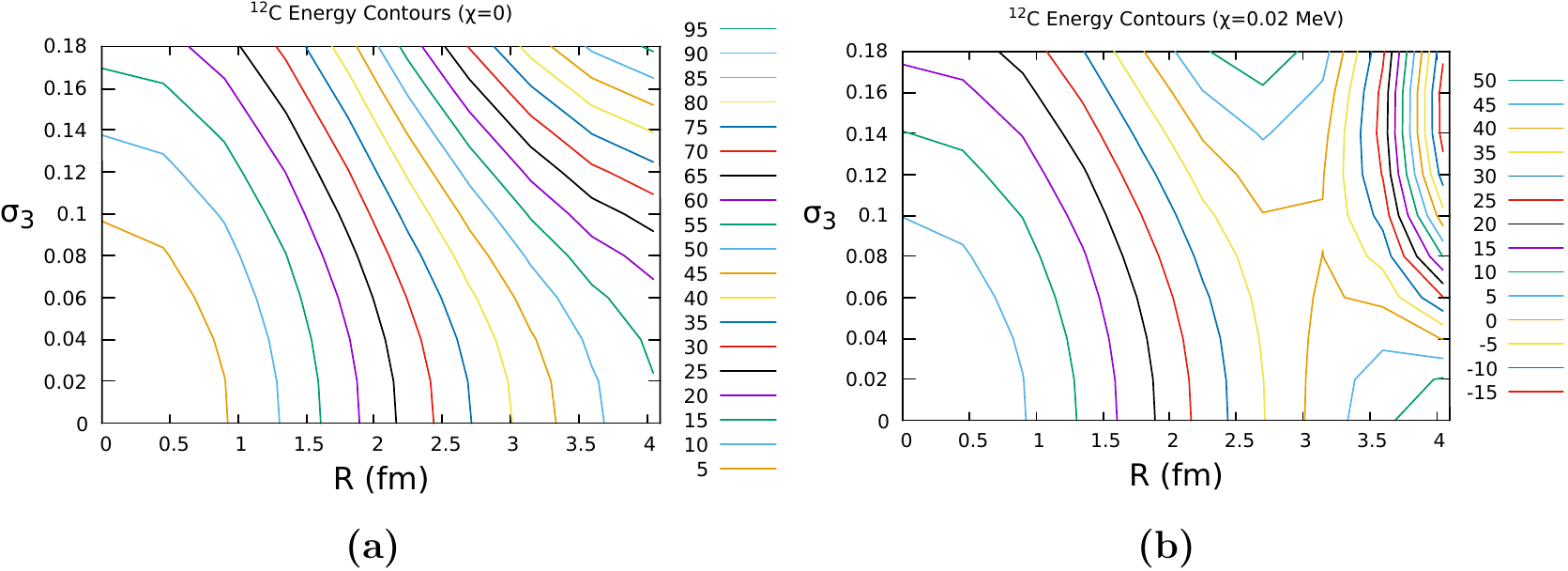}
\centering
\caption{(color online). The energy contour  of $^{12}$C on the $(R,\sigma_3)$ plane where the parameters $(d,a_2)$  have been chosen to minimize
the energy for each $R$ at $\sigma_3$=0.  Figure (a) gives the results  without the octupole-octupole interaction ( $\chi=0$), and figure (b) the results  for  the octupole-octupole interaction $\chi$=0.02 MeV in Eq. (\ref{chi}).
}
\label{twoa3}
\end{figure}

Upon fixing the $(d,a_2)$ values to be those that give an energy  minimum for the case of $\chi=0$, we  study then the variations of the energy surface on the 
$(R, \sigma_3)$ plane.  In Fig.\ \ref{twoa3}(a)  for $\chi=0$
without the octupole-octupole interaction, the energy surface at the ground state at the ground state ($R$=0) is an energy minimum also in $\sigma_3$.    At the Hoyle energy ($R$=1.16 fm), the energy surface rises as a function of $\sigma_3$.  
In Fig.\ \ref{twoa3}(b) for $\chi$=0.02 MeV
with  a small octupole-octupole interaction strength, the energy surface has a saddle point at $(R=3 {\rm ~fm},\sigma_3\sim 0.08)$ with  a barrier height of about 35-40 MeV.   Upon passing over the saddle point beyond $R$$\gtrsim$ 3 fm, the energy surface drops down which reflects the three alpha cluster structure, indicating that there is a high barrier separating the region of the ground state energy minimum and the region of the three-alpha cluster.  The general features of the energy surface landscape for other values of $\chi$ are similar to those in Fig.\ \ref{twoa3}(b).
At greater values of $\chi$, the barrier moves lower at a slightly lower value of $R$. 
 There is however no energy minimum at the Hoyle energy of $E_x$=7.654 MeV for 
a linear  residual octupole-octupole interaction as given in Eq.\ (\ref{chi}).

The result indicates that with only a single Slater determinant in the expansion of the generator coordinate sum in Eq. (\ref{mix2}) in the toroidal mean-field model and the linear 
octupole-octupole interaction,
there is no energy minimum at the Hoyle energy of $E_x$=7.654 MeV.  
The search for a local energy minimum at the Hoyle energy will likely require
either a quadratic  octupole-octupole interaction or the  inclusion of many more Slater determinants.

\section{Conclusions and Discussions}

We study the intrinsic  oblate states of $^{12}$C from the generator coordinate viewpoints.  We note that under the assumption of planar dominance in which the intrinsic motion of the nucleons is confined in a planar region for the nucleus, the Griffin-Hill-Wheeler equation can be substantially simplified.  It is only necessary to solve the equation in the planar region, and the solution of the intrinsic state with quantum number $K$ can be projected out to obtain the rotational band of states with angular momentum $I$ and component $M$ in the laboratory frame.
For a basic shape in the form of a triangular cluster of three alpha particles under the planar dominance,
the generator coordinate sum of the orientations of the cluster triangle 
 lead to toroidal density distributions.

One can take the alternative generator coordinate sum whose elements are toroidal wave functions  
 and obtain a similar Griffin-Hill-Wheeler equation.  The case of a single Slater determinant  then leads us to the toroidal mean field approximation.

It turns out that a single-determinant mean-field approximation is adequate to describe  the 
ground state of $^{12}$C which gives a toroidal core immersed in oblate spheroids.  The mean field approximation also gives  a toroidal density at the Hoyle state, but the energy of the state lies on a slope as a function of the radial parameter and is unstable with respect to the radial contraction.  
It shows that to study the stability of  the Hoyle state, it  will be necessary to go beyond the mean field.  

We are however motivated to continue the search for a local toroidal energy minimum for the Hoyle state because  phenomenological comparison of the spectrum with  the  toroidal nucleus signature suggests that 
the Hoyle state and many
of its higher excited states may be tentatively attributed to those of
the $^{12}$C nucleus in a toroidal configuration.
Our next task is to  introduce a two-body residual  interaction and use the toroidal  mean field or the toroidal shell model basis in the Griffin-Hill-Wheeler equation  to study the Hoyle state and its associated excited states.
  Results from such calculations will be of great interest.

\section*{Acknowledgements}

The author thank Profs.\ J. J. Griffin,  H. Feldmeier, A. Jensen, P. Ring, Jie Meng, Peng-Wei Zhao, Qi-Bo Chen,   A. Staszczak,  S. Aberg, P. Quentin,  L. Bonneau for
helpful communications and discussions.  The research was supported in
part by the Division of Nuclear Physics, U.S. Department of Energy
under Contract DE-AC05-00OR22725.

\end{document}